\documentclass[aps,reprint,superscriptaddress,longbibliography,prb,floatfix]{revtex4-2}
\usepackage{graphicx}
\usepackage{amsmath}

\usepackage{amsfonts}
\usepackage{comment}
\usepackage{bbm}
\usepackage{bm}
\usepackage{bbold}
\usepackage{comment}
\usepackage{esint}
\usepackage{cancel}
\usepackage[dvipsnames]{xcolor}
\usepackage{microtype}
\usepackage[normalem]{ulem}
\usepackage{mathrsfs}
\usepackage{multirow}
\usepackage{array}
\usepackage{lipsum}
\usepackage{svg}
\usepackage{systeme}
\usepackage{dutchcal}
\usepackage{tikz}
\usepackage{pgfplots}
\pgfplotsset{compat=1.18}
\usepackage[unicode=true, colorlinks=true, citecolor={blue!80!black}, urlcolor={blue!50!black}, linkcolor = {blue!80!black}]{hyperref}

\newcommand{\abs}[1]{\left\vert#1\right\vert}
\newcommand{\ket}[1]{\left\vert#1\right\rangle}
\newcommand{\bra}[1]{\left\langle#1\right\vert}
\newcommand{\braket}[2]{\left\langle#1\vert#2\right\rangle}

\graphicspath{{../figures/}}

\begin{document}
\title{
Chiral topological phases of proximitized planar germanium
}
\author{Vittorio Coppini}
\affiliation{Dipartimento di Fisica, Sapienza Università di Roma, Piazzale Aldo Moro 2, 00185 Rome, Italy}
\author{Bernard van Heck}
\affiliation{Dipartimento di Fisica, Sapienza Università di Roma, Piazzale Aldo Moro 2, 00185 Rome, Italy}

\date{\today}
\begin{abstract}
We investigate the topological phase diagram of a two-dimensional hole gas in planar germanium resulting from the combination of spin-orbit coupling, proximity-induced $s$-wave superconductivity, and magnetization.
The peculiarity of the hole gas is the presence of a Rashba spin-orbit interaction cubic in momentum, possibly coexisting with a linear term, which results in the presence of distinct topological phases with Chern numbers as high as four. 
We map out the topological phase diagram based on effective single-band models with and without spin, and use input parameters based on a self-consistent Schrödinger-Poisson calculation of a planar germanium hole gas.
As a concrete model for inducing magnetization in the hole gas, we consider the exchange interaction coming from the presence of a dilute ensemble of manganese impurities.
A topological phase with $C=3$ could be accessible experimentally in unstrained planar germanium.
\end{abstract}

\maketitle

%\tableofcontents

\section{Introduction}

\subsection{Motivation}

Two-dimensional superconducting systems with broken time-reversal symmetry have attracted the interest of condensed matter physicists since the seminal work of Read and Green~\cite{read2000}.
Belonging to class D of the Altland-Zirnbauer classification~\cite{altland1997}, in the periodic table of topological insulators and superconductors~\cite{kitaev2009,ryu2010} they are assigned a $\mathbb{Z}$ topological invariant, the Chern number~$C$. 

By the bulk-boundary correspondence, a non-zero value of $C$ protects the presence of chiral edge modes along the edges of finite samples or at the boundaries of domains with different values of $C$. Furthermore, for odd values of $C$ vortices bind zero-energy Majorana modes with non-Abelian exchange statistics~\cite{ivanov2001}. Chiral edge modes and vortices are linked with interesting transport signatures~\cite{akhmerov2009} and topological quantum computation~\cite{nayak2008}.

So far it has not been easy to find a material that spontaneously exhibits the triplet pairing required to observe chiral superconductivity.
For quite a while Sr$_2$RuO$_4$ was a possible candidate~\cite{mackenzie2003}, but chiral $p$-wave superconductivity has now been ruled out~\cite{pustogow2019}.
Other, more recent, exciting possibilities are UTe$_2$~\cite{jiao2020} and graphene multilayers~\cite{han2025}.

A parallel route to obtain these chiral topological phases has been the material engineering route, primarily based on the hybridization of two-dimensional systems with conventional $s$-wave superconductors. The list of proposals falling under this umbrella is long~\cite{alicea2012,flensberg2021}. 
Possible starting points are the surface states of topological insulators~\cite{fu2008}, the quantum Hall or quantum anomalous Hall state~\cite{qi2010}, and two-dimensional electron gases in semiconductors with strong spin-orbit coupling~\cite{sau2010,alicea2010,sato2010,sau2010-2}.

In this vein, Ref.~\cite{mao2011} studied the proximity effect in hole-doped semiconductors such as Ge. The main difference with respect to 2D electron gases in semiconductors such as InAs is that, in the presence of strong confinement breaking the inversion symmetry, the spin-orbit interaction of the 2D hole gas is cubic rather than linear in momentum. This yields a chiral topological phase with $C=3$, due to an effective $f$-wave pairing.

Recently, there has been a surge of interest in the 2D hole gas in germanium-based heterostructures as a material platform for quantum applications such as spin qubits~\cite{scappucci2021}.
The proximity effect from superconducting materials and alloys has also been reported~\cite{hendrickx2018,aggarwal2021,tosato2023}, indicating that pairing can be induced by proximity in the hole gas.
Furthermore, the high mobilities of Ge/SiGe heterostructures~\cite{dobbie2012,sammak2019} are an attractive starting point to study superconducting phases that, because time-reversal symmetry is broken, are not protected by Anderson’s theorem~\cite{anderson1959} and are therefore sensitive to disorder.

Given these developments, we find it timely to study in greater detail the topological phase diagram resulting from the proximitization of a 2D hole gas by an $s$-wave superconductor. 

\subsection{Overview of models and results}

We tackle the description of topological phases in planar germanium using three models of increasing complexity: two toy models plus an effective low-energy model.
We begin in Sec.~\ref{sec:model1} by studying a spinless toy model, an extension of the Read-Green model which includes a pairing term which is cubic in momentum.
The model features a topological phase transition between two topological phases with $C=-1$ and $C=3$, depending on the relative strength of the linear and cubic pairing terms.
We verify the bulk-edge correspondence and point out that due to cubic distortion of the dispersion relation a counter-propagating edge mode can be present in the $C=-1$ phase.

In Sec.~\ref{sec:model2} we consider a spinful single-band toy model, which includes a spin-orbit interaction with linear and cubic terms in momentum, an effective Zeeman field, and $s$-wave superconducting pairing.
This model can be seen as the appropriate extension to the germanium hole gas of well-studied models of topological superconductivity in 2D electron gases of other semiconducting platforms such as InAs \cite{sau2010}.
The topological phase diagram features phases with $\abs{C}=1, 3$ and $4$, and three distinct critical lines that can be traced back to the zeros of the spin-orbit interaction term.
The occurrence of Chern numbers as high as four can be explained via a reduction of the spinful model to two copies of the spinless model, with pairing terms of opposite chirality.

\begin{figure}
    \centering
    \includegraphics[width=0.9\linewidth]{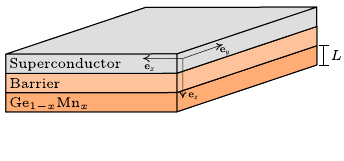}
    \caption{A two-dimensional germanium heterostructure suitable to realize chiral topological phases. An unstrained germanium layer is lightly doped with manganese impurities. The concentration $x$ indicates the fraction of Mn atoms: relevant values for this work are well below $x=0.01$. The germanium hole gas is weakly coupled to an $s$-wave superconductor deposited on the top side, via a Ge-heavy barrier layer such as Si$_{0.2}$Ge$_{0.8}$.}
    \label{fig:sketch}
\end{figure}

In Sec.~\ref{sec:model3}, we model more in detail the heterostructure shown in Fig.~\ref{fig:sketch}, starting from a microscopic description of the hole gas based on a six-band Luttinger-Kohn model with confinement.
As a concrete model to induce time-reversal symmetry breaking in the hole gas, we include the presence of a diluted ensemble of manganese impurities, which polarizes the germanium holes via the exchange interaction~\cite{dietl2001}.
We derive an effective planar, single-band Hamiltonian for this system: it takes a form very similar to that of the toy model of Sec.~\ref{sec:model2}, with the addition of anisotropic effects due to the coupling between spin and momentum induced by the magnetization.
Using values for the model parameters extracted from a self-consistent Schrödinger-Poisson calculation, we study the phase diagram of the superconducting system for both out-of-plane and in-plane magnetization.
In the former case, a topological phase with $C=3$ emerges at a fairly low density of magnetic impurities.
The latter case is characterized by the lack of gapped topological phases and the anisotropy of the magnetic response.

\subsection{Modeling assumptions}

We list here the main assumptions that enter our models.
First, we assume that singlet superconducting pairing can be induced in the hole gas by the proximity to an $s$-wave superconductor.
We include the pairing in the simplest possible way, as a constant pairing potential $\Delta_0$, thus neglecting the possible frequency and spatial dependence that would come from a more realistic treatment of the self-energy term induced by a neighboring superconductor.
This simple treatment is appropriate for a weak coupling regime, with the proximity effect mediated by a barrier layer separating the parent superconductor and the germanium hole gas.

Recent works have highlighted that a microscopic model of the proximity effect between an $s$-wave superconductor and the germanium holes can generate momentum-dependent pairing terms as well~\cite{babkin2025,pino2025}.
As a starting point, we stick to the simplest $s$-wave pairing because we expect it to be the most robust with respect to scattering and disorder at the interface in a real sample.
Furthermore, all our calculations apply to a regime in which a single two-dimensional sub-band is occupied, and so multi-band pairing terms can be neglected.

A requirement to induce a chiral topological phase in the hole gas is to break time-reversal symmetry while preserving the superconducting pairing.
Furthermore, because the spin-orbit interaction locks the spin to the momentum in the plane, in order to induce chiral topological phases in the hole gas it is necessary to polarize their spin out of the plane.
This makes it impractical to use an external magnetic field to break time-reversal symmetry, since it would have to be applied perpendicular to the sample and thus it would likely be incompatible with preserving superconductivity in the parent superconductor.
At the theoretical level, one can imagine that the time-reversal symmetry breaking comes from the proximity to a ferromagnetic insulator, which produces an effective Zeeman field, as was indeed imagined from the very early proposals~\cite{sau2010}.
On the material side, such a solution has been tested by introducing layers of the ferromagnetic material EuS in hybrid heterostructures based on InAs~\cite{vaitiekenas2021,escribano2022}, but never, to the best of our knowledge, in germanium.

Therefore, as a concrete solution originating from the literature of dilute ferromagnetic semiconductors~\cite{dietl2014}, we consider a germanium layer doped with dilute manganese impurities, leading to an internal Zeeman field originating from the exchange interaction between the $p$-type valence bands of germanium and $d$-orbitals of Mn atoms.
To this end, in Sec.~\ref{sec:model3} we adopt the exchange interaction Hamiltonian derived in Ref.~\cite{dietl2001}, with the caveat that we apply it to a regime of much lower Mn content and Ge hole densities.
Indeed, the Mn atoms in Ge are deep acceptors~\cite{woodbury1955}, and so at the low temperatures relevant for the superconducting phase diagram they are frozen out and do not dope the germanium valence band with holes.
We will therefore assume the hole density to be defined by a shallow quantum well at the germanium-barrier interface; in our self-consistent calculations it will be of the order of $10^{10}$~cm$^{-2}$.
We will then see in Sec.~\ref{sec:model3} that Mn concentrations as low as $0.1\%$ are sufficient to drive a topological transition in the hole gas.

At these low Mn concentrations and hole densities, the ferromagnetic ordering studied in Ref.~\cite{dietl2001} may not occur, or it may occur with much smaller Curie temperatures.
Even in this case, however, a large internal Zeeman field may be activated by a small external magnetic field, exploiting the paramagnetic response of the Mn impurities.
Apart from the small field possibly applied to polarize the impurity, the spin splitting originates from exchange rather than an applied field, and the stray field originating from the resulting magnetization is small.
Thus orbital effects, which are usually detrimental to superconducting topological phases in class $D$~\cite{nijholt2016}, can be neglected and we do not include any vector potential in the Hamiltonian.

Because the magnetization density is treated at the mean field level, averaging out the fluctuations that would arise in a single ensemble realization, we can work with a clean model which preserves the translation symmetries in the plane.
This is despite the fact that the introduction of an ensemble of manganese impurities should be expected to degrade the high mobility that can be reached in germanium hole gases.
The addition of disorder into the model would certainly compromise the energy gap in the topological phases with broken time-reversal symmetry, possibly leading to the closure of the gap and a transition to a disordered phase of gapless superconductivity.
We return to this issue in the concluding remarks.

\subsection{Relation to previous research}

We mention here previous literature that has informed our work.
On the experimental side, we have been inspired by results demonstrating the feasibility of growing planar germanium in combination with different superconducting materials in recent years~\cite{hendrickx2018,aggarwal2021,tosato2023,strohbeen2024,valentini2024,steele2025,fabris2026}.

Our toy models reproduce the general strategy of obtaining topological phases of matter by the design of suitable semiconductor-superconductor heterostructures, which was pioneered in several works exploiting the Rashba spin-orbit interaction in 2D electron gases~\cite{sato2009,sau2010,alicea2010,sato2010,sau2010-2}, followed up by experimental progress~\cite{shabani2016,kjaergaard2016,phan2022}.

As mentioned in the introduction, this idea has already been extended to hole-based semiconductors~\cite{mao2011}.
In particular, several works have studied topological superconducting phases hosting Majorana zero modes in germanium-based wires of different kinds~\cite{maier2014,adelsberger2023,laubscher2024_2,laubscher2024}.
Our work can be considered as the counterpart focusing on planar topological phases, extending previous studies of two-dimensional superconducting models with different powers of momentum involved in the pairing~\cite{mao2011,shitade2016,lepori2018}.

The spin-orbit interaction in planar germanium has been the subject of several studies, given its importance for spin qubit applications \cite{scappucci2021,morita2014,morrison2014,failla2015}. Theoretically, the possible co-existence of linear and cubic terms, their relative sign and magnitudes, and the crucial role of strain have all been studied~\cite{winkler2003spin,durnev2014,Marcellina2017,terrazos2021,xiong2021,bosco2021}.
The Schrieffer-Wolff calculations reported in this work reproduce the results of Ref.~\cite{Marcellina2017}, extending them with the inclusion of the magnetic terms arising from the $p$-$d$ orbital exchange interaction in the six-band Luttinger-Kohn model. Importantly, we derive our effective parameters using a self-consistent treatment of the confining potential, and in the relevant regime of low hole densities.

\section{Spinless toy model}
\label{sec:model1}

\begin{figure*}[t!]
    \centering
    \includegraphics[width=\textwidth]{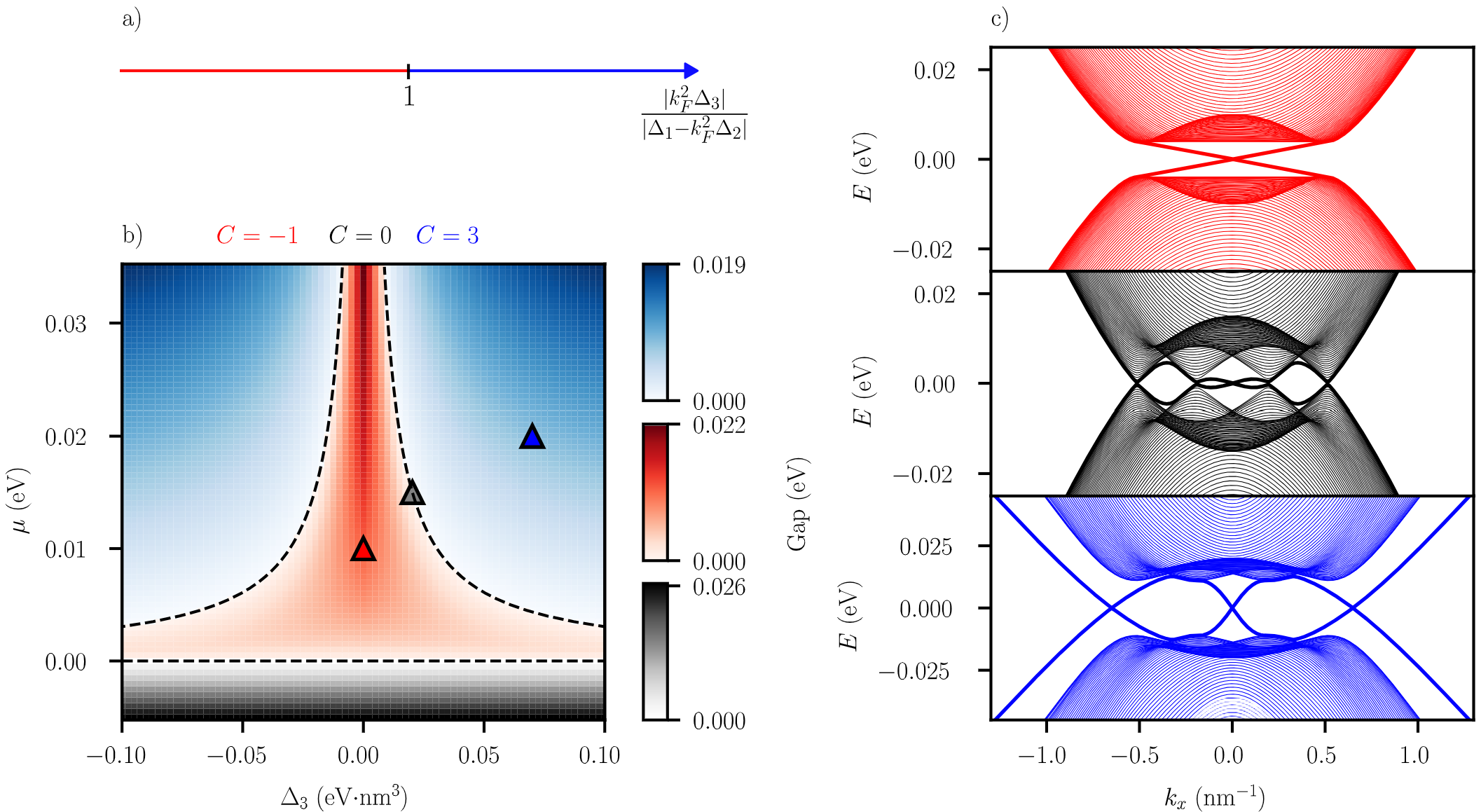}
    \caption{Spinless model. (a) Projected topological phase diagram. 1 is the critical point where the transition occurs. (b) Example of topological phase diagram for representative values of the parameters $\Delta$ ($\Delta_1 = 0.008$ eV$\cdot$nm and $\Delta_2=0$ eV$\cdot$nm$^3$). The dashed hyperbolas represent the critical lines in Eq. \ref{eq:model_one_critical_line_one} and \ref{eq:model_one_critical_line_two}, while the horizontal line is the critical line $\mu=0$. The triangular markers indicate the values of $\mu$ and $\Delta_3$ used in panel (c). (c) One-dimensional band structure of an infinite lead ($W=200$ nm, $\Delta_1 = 0.008$ eV$\cdot$nm and $\Delta_2=0$ eV$\cdot$nm$^3$) for the corresponding parameters values of the points marked by a triangle in panel (b). Zero energy states are clearly visible, signaling the presence of edge states.}
    \label{fig:model_one}
\end{figure*}

The first toy model we consider is an extension of the Read-Green model that includes pairing terms which are cubic in momentum.
It is defined by the following Hamiltonian for spinless fermions of mass $m$ in two dimensions:
\begin{equation}
    H=\int d\mathbf{k}\,\left[\zeta_\mathbf{k}\,c^\dagger_\mathbf{k} c_\mathbf{k} +\tfrac{1}{2}\Delta^*_\mathbf{k}\,c^\dagger_\mathbf{k} c^\dagger_{-\mathbf{k}}+\textrm{h.c.}\right]\,.
\end{equation}
Here,
\begin{equation}\label{eq:zeta_k}
\zeta_\mathbf{k} = \frac{\hbar^2k^2}{2m}-\mu\,,
\end{equation}
$\mu$ is the chemical potential, and
\begin{equation}
\label{eq:pairing_model_one}
\Delta_\mathbf{k} = (\Delta_1 - \Delta_2 k^2)\, (k_x+ik_y) + \Delta_3(k_x-ik_y)^3\,.
\end{equation}
is the superconducting pairing, characterized by three real parameters $\Delta_1$, $\Delta_2$ and $\Delta_3$.
This choice for the pairing is motivated by the form of the spin-orbit interaction in the germanium hole gas, which will be discussed in more detail in later Sections.
Like the latter, it can also be justified on the basis of symmetry.
Here, we note that if only one of the two terms $\sim(k_x+ik_y)$ and $\sim(k_x-ik_y)^3$ in Eq.~\eqref{eq:pairing_model_one} is non-zero, the Hamiltonian is invariant under rotations in the plane.
On the other hand, if both terms are present, the symmetry group is reduced to $C_4$, compatible with the symmetry group of crystalline planar germanium.

By introducing the Nambu spinor
\begin{equation}
\psi_\mathbf{k}=\begin{bmatrix}
c_\mathbf{k} \\ c^\dagger_{-\mathbf{k}}
\end{bmatrix}\,,
\end{equation}
up to a constant we can rewrite the Hamiltonian as
\begin{equation}
H =\frac{1}{2}\int d\mathbf{k}\,\psi^\dagger_\mathbf{k}\, h_\mathbf{k}\,\psi_\mathbf{k}\,,
\end{equation}
with a Bogoliubov-de Gennes Hamiltonian
\begin{equation}
h_\mathbf{k} = \begin{bmatrix}
\zeta_\mathbf{k} & \Delta^*_\mathbf{k} \\
\Delta_\mathbf{k} & -\zeta_\mathbf{k}
\end{bmatrix}\,.
\end{equation}
The single-particle excitation energies are given by the eigenvalues of $h_\mathbf{k}$, namely
\begin{equation}
E_\mathbf{k} = \sqrt{\zeta_\mathbf{k}^2+\abs{\Delta_{\mathbf{k}}}^2}\,.
\end{equation}
The energy gap may close only if $\zeta_\mathbf{k}=0$ and $\Delta_\mathbf{k}=0$ simultaneously.
There are two distinct scenarios where this may happen.

The first scenario takes place if $\mathbf{k}=0$ and $\mu=0$: it signals a topological transition from the trivial phase with $C=0$ when $\mu<0$ to a topological superconductor with a finite value of $C$ when $\mu>0$.
This is the original topological transition of the Read-Green model, except that in this case the Chern number on the topological side of the transition can take two different values.
We can see this by considering some limiting cases.
On one hand, if $\Delta_3=0$ so that only the term $\sim(k_x+ik_y)$ is present in the Hamiltonian, then $C=-1$.
On the other hand, if $\Delta_1=\Delta_2=0$ so that only the term $\sim(k_x-ik_y)^3$ is present, then $C=3$.

The second scenario represents a distinct topological phase transition that separates the two limiting cases just mentioned. 
It occurs at $k=k_F$, where $k_F=(2m\mu/\hbar^2)^{1/2}$ is the Fermi quasi-momentum, provided that the condition
\begin{equation}\label{eq:gap_closing_condition_model_one}
(\Delta_1-k_F^2\Delta_2)=-e^{-4i\theta_\mathbf{k}}k_F^2\,\Delta_3
\end{equation}
is also satisfied, which originates from the requirement $\Delta_\mathbf{k}=0$, and where we introduced $\theta_\mathbf{k}=\arg(k_x+ik_y)$.

When the magnitude of right hand side in Eq.~\eqref{eq:gap_closing_condition_model_one} is larger than that of the left hand side, the $\sim(k_x-ik_y)^3$ term dominates the pairing and the system is on the $C=3$ side of the transition.
When the converse is true, the $\sim(k_x+ik_y)$ term dominates and the system is on the $C=-1$ side of the transition.
We thus obtain the schematic phase diagram of Fig.~\ref{fig:model_one}(a).

The condition in~Eq.~\eqref{eq:gap_closing_condition_model_one} yields two curves in parameter space at which the transition can occur, depending on whether one chooses $\exp(4i\theta_\mathbf{k})=\pm 1$. With the minus sign, the gap closes on the main diagonals in momentum space and when
\begin{equation}\label{eq:model_one_critical_line_one}
\mu = \frac{\hbar^2}{2m}\frac{\Delta_1}{\Delta_2+\Delta_3}\,.
\end{equation}
With the plus sign, the gap closes on the principal axes in momentum space and when
\begin{equation}\label{eq:model_one_critical_line_two}
\mu = \frac{\hbar^2}{2m}\frac{\Delta_1}{\Delta_2-\Delta_3}\,,
\end{equation}
provided the right hand side is positive.

These two critical lines give rise to the topological phase diagram shown in Fig.~\ref{fig:model_one}(b) for the case $\Delta_2=0$.
We note that along these critical lines, the Chern number changes by four, consistent with the closing and reopening of the gap at four distinct Dirac points in momentum space.

A notable and reasonable feature of the phase diagram is that the topological phase with $C=-1$ is always favored at small chemical potentials, because the linear pairing dominates the cubic pairing at small momenta. Thus a direct transition from $C=0$ to $C=3$ is only possible if the linear pairing term is absent ($\Delta_1=0$).

In Fig.~\ref{fig:model_one}(c) we verify the bulk-boundary correspondence by inspecting the band structure of translationally-invariant, finite-width lead oriented along the $x$ direction at representative points in the topological phase diagram. 
In this lead geometry, at energies below the bulk gap, we expect the one-dimensional band-structure to display left- or right-moving chiral edge modes localized at opposite edges of the sample.

And indeed, in the upper panel of Fig.~\ref{fig:model_one}(c), corresponding to the $C=-1$ phase, we see one pair of counter-propagating modes linearly dispersing around $k=0$.
In the lower panel of Fig.~\ref{fig:model_one}(c), corresponding to the $C=3$ phase, we see instead three pairs of counter-propagating modes, around $k_x=0$ and $k_x\approx \pm k_F$.
The middle panel displays the one-dimensional band-structure right on the critical line~\eqref{eq:model_one_critical_line_one}: the one-dimensional projection of the four bulk Dirac cones is clearly visible. In between the two inner Dirac cones, one pair of counter-propagating edge modes centered around $k_x=0$ remains.

\begin{figure}
    \centering
    \includegraphics[width=\linewidth]{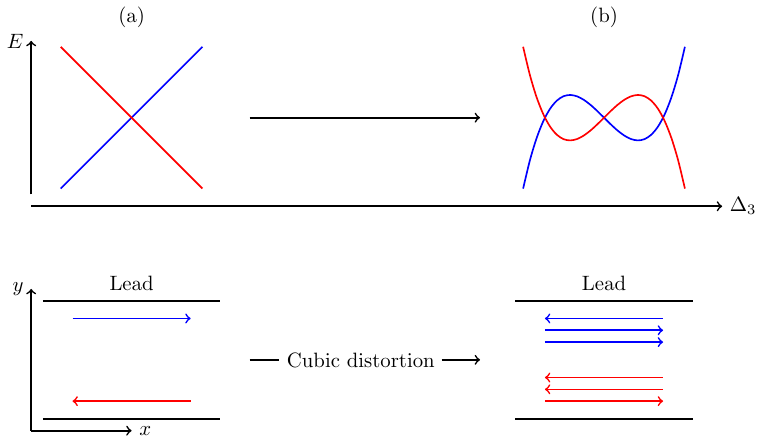}
    \caption{Sketch of the distortion of the edge modes dispersion for increasing cubic term $\Delta_3$, within the $C=-1$ phase. (a) When $\Delta_3=0$ the dispersion of the chiral modes is linear (top), and there is a single propagating chiral mode along one edge of the sample (bottom). 
    (b) When $\Delta_3$ dominates over the linear term, the dispersion of the chiral modes is distorted (top), and features three zero energy crossings. There are now three edge modes per edge of the sample, including one upstream mode (bottom). The change from panels (a) to (b) does not require a topological phase transition, and the Chern number is conserved.}
    \label{fig:linear2cubic}
\end{figure}

The dispersion of the chiral edge modes in the $C=-1$ phase is interesting.
When $\Delta_3=0$, the dispersion of the pair of chiral modes is purely linear, as sketched in Fig. \ref{fig:linear2cubic} (a), but when $\Delta_3$ is finite it acquires a cubic distortion.
When the cubic part of the dispersion of the edge modes becomes large enough, the number of zero-energy crossings changes from one to three, \emph{before} the bulk gap closes at the topological transition.
From this point onward, there are \emph{three} propagating modes along a single edge, but one of them is an upstream mode, consistently with the Chern number $C=-1$, as sketched in Fig. \ref{fig:linear2cubic}.
Finally, with the gap closing at the topological transition, the edge states re-arrange to feature three co-propagating modes.

\section{Spinful single-band toy model}
\label{sec:model2}

We now move on to the study of a single-band toy model.
Similar to the strategy in the previous section, we generalize the well-known model of topological superconductivity based on a Rashba 2DEG to a model with a spin-orbit interaction which has cubic terms.
The model is:
\begin{equation}\label{eq:H_model2_integral}
H = \int d\mathbf{k} \left[C^\dagger_\mathbf{k}\mathcal{h}_kC_\mathbf{k} +\Delta_0(c^\dagger_{\mathbf{k}\uparrow}c^\dagger_{-\mathbf{k}\downarrow}+c_{-\mathbf{k}\downarrow}c_{\mathbf{k}\uparrow})\right]\,,
\end{equation}
where $\Delta_0$ is a proximity-induced $s$-wave pairing gap,
\begin{equation}
C_\mathbf{k} = \begin{bmatrix} c_{\mathbf{k}\uparrow} \\ c_{\mathbf{k}\downarrow}
\end{bmatrix}\,,
\end{equation}
and $\mathcal{h}_\mathbf{k}$ the following normal-state, single-particle Hamiltonian:
\begin{equation}\label{eq:hk_singleband}
\mathcal{h}_\mathbf{k} = \zeta_\mathbf{k}\,\mathbb{I}+\boldsymbol{\sigma}\cdot\mathbf{M}+A_\mathbf{k}\sigma_+ + A^*_{\mathbf{k}}\sigma_-\,.
\end{equation}
Here, $\zeta_\mathbf{k}$ is the same as in Eq.~\eqref{eq:zeta_k}, $\mathbb{I}$ is the $2\times 2$ identity matrix, $\boldsymbol{\sigma}$ is a vector of Pauli matrices, and $\mathbf{M}$ is an external magnetization coupling to the spin, due e.g.~to ordered magnetic moments in the heterostructure. Finally,
\begin{equation}\label{eq:spin_orbit_modello_due}
A_\mathbf{k} = \frac{i\hbar^2}{2m}[(\alpha_1-\alpha_2 k^2) (k_x+ik_y) + \alpha_3 (k_x-ik_y)^3]
\end{equation}
is the spin-orbit interaction, which comprises terms both linear and cubic in momentum.

This form of the spin-orbit interaction can be understood considering how the symmetry group changes when we lower the dimensionality of the system. 
If we confine bulk germanium holes with an inversion-asymmetric potential (necessary for having finite spin-orbit interaction), the dimensional reduction to a plane reduces the symmetry group from the cubic group $O_h$ to $C_{4v}$~\cite{winkler2003spin}~\footnote{The addition of a magnetization term along $\hat{z}$ further reduces $C_{4v}$ to the group $C_4$ relevant for Sec.~\ref{sec:model2}}. 
Therefore, the confined system is symmetric under $C_4$ rotations around the axis perpendicular to the plane and so must be $A_\mathbf{k}$.
The values of the parameters $\alpha_1, \alpha_2$ and $\alpha_3$ can be obtained from a Schrieffer-Wolff perturbative expansion of the Luttinger-Kohn Hamiltonian of confined germanium around $\mathbf{k}=\mathbf{0}$, as will be done in the next Section.
For the time being, we take $\alpha_1, \alpha_2, \alpha_3$ to be independent \emph{positive} parameters.
Note that with the parametrization adopted in Eq.~\eqref{eq:spin_orbit_modello_due}, the coefficient $\alpha_1$ has the dimensions of an inverse length, while $\alpha_2$ and $\alpha_3$ have the dimensions of a length.

\subsection{Normal state properties}

Before studying the superconducting phase diagram of the model, it is useful to briefly review the normal state properties associated with the Hamiltonian $\mathcal{h}_\mathbf{k}$. Its eigenvalues are given by 
\begin{equation}
E_{\mathbf{k},\pm} = \zeta_\mathbf{k}\pm \sqrt{M_z^2 + (M_x+X_\mathbf{k})^2 + (M_y + Y_\mathbf{k})^2}\,,
\end{equation}
where we decomposed $A_\mathbf{k}$ into its real and imaginary parts, $A_\mathbf{k}=X_\mathbf{k}-i Y_\mathbf{k}$.

When $\mathbf{M}=\mathbf{0}$, the expression simplifies to
\begin{equation}
E_{\mathbf{k},\pm} = \zeta_\mathbf{k}\pm\abs{A_\mathbf{k}}.
\end{equation}
Since
\begin{equation}\label{eq:A_mod}
\begin{split}
\abs{A_\mathbf{k}}=k &\left[(\alpha_1-\alpha_2 k^2)^2+\alpha_3^2 k^4\right.\\ &\left.+2\alpha_3k^2(\alpha_1-\alpha_2 k^2)\cos(4\theta_\mathbf{k})\right]^{1/2}\,,
\end{split}
\end{equation}
we see that, due to the spin-orbit coupling, the energy bands are spin-split except at special points in momentum space where $A_\mathbf{k}=0$.
Given that $A_\mathbf{k}$ has the same dependence on $\mathbf{k}$ as $\Delta_\mathbf{k}$ of Eq.~\eqref{eq:pairing_model_one}, these points are the same as the ones discussed in the previous Section.
Namely: (i) the origin $k=0$; (ii) four points on the main diagonals in momentum space and magnitude $k_0$ with
\begin{equation}\label{eq:k0}
k_0^2 = \frac{\alpha_1}{\alpha_2+\alpha_3}\,,
\end{equation}
(iii) if $\alpha_2>\alpha_3$, four points on the principal axes and magnitude $k_1$ with
\begin{equation}\label{eq:k1}
k_1^2 = \frac{\alpha_1}{\alpha_2-\alpha_3}\,,
\end{equation}
and finally, (iv) only if $\alpha_3=0$, the circumference in momentum space with radius $(\alpha_1/\alpha_2)^{1/2}$.

Also similar to the previous Section is the fact that the energy bands are anisotropic if both terms are present in Eq.~\eqref{eq:spin_orbit_modello_due}: their simultaneous presence reduces the rotational symmetry to a four-fold rotation symmetry around the origin in momentum space. The anisotropy is due to the last term of Eq.~\eqref{eq:A_mod} and it is proportional to the product $\alpha_3 (\alpha_1-\alpha_2 k^2)$.

The qualitative effect of $\mathbf{M}$ on the energy bands is the following. An out-of-plane magnetization opens an avoided crossing at the spin-degenerate points listed above.
It creates a Zeeman gap between the bands while maintaining the four-fold rotation symmetry.
On the other hand, an in-plane magnetization removes the four-fold rotation symmetry by tilting the energy bands along its direction.
It also opens an avoided crossing at the spin-degeneracy points listed above, except if they occur along a direction in momentum space aligned with $\mathbf{M}$.
In this case, the degeneracy points ``slide'' along the energy bands until the magnetization is so large that the energy bands do not cross anymore.

Since the form of the spin-orbit interaction~\eqref{eq:spin_orbit_modello_due} is the result of an expansion at small momenta, the model has a finite range of validity in momentum space.
Indeed, the model ceases to be valid when the cubic terms in the spin-orbit interaction become comparable to the quadratic kinetic energy.
After this point, the energy of the lower energy band $E_{\mathbf{k},-}$ starts decreasing with increasing $k$, and ceases being physical.
In numerical calculations, we always check whether the chemical potential is low enough to avoid this regime.

\begin{figure*}[t!]
    \centering
    \includegraphics{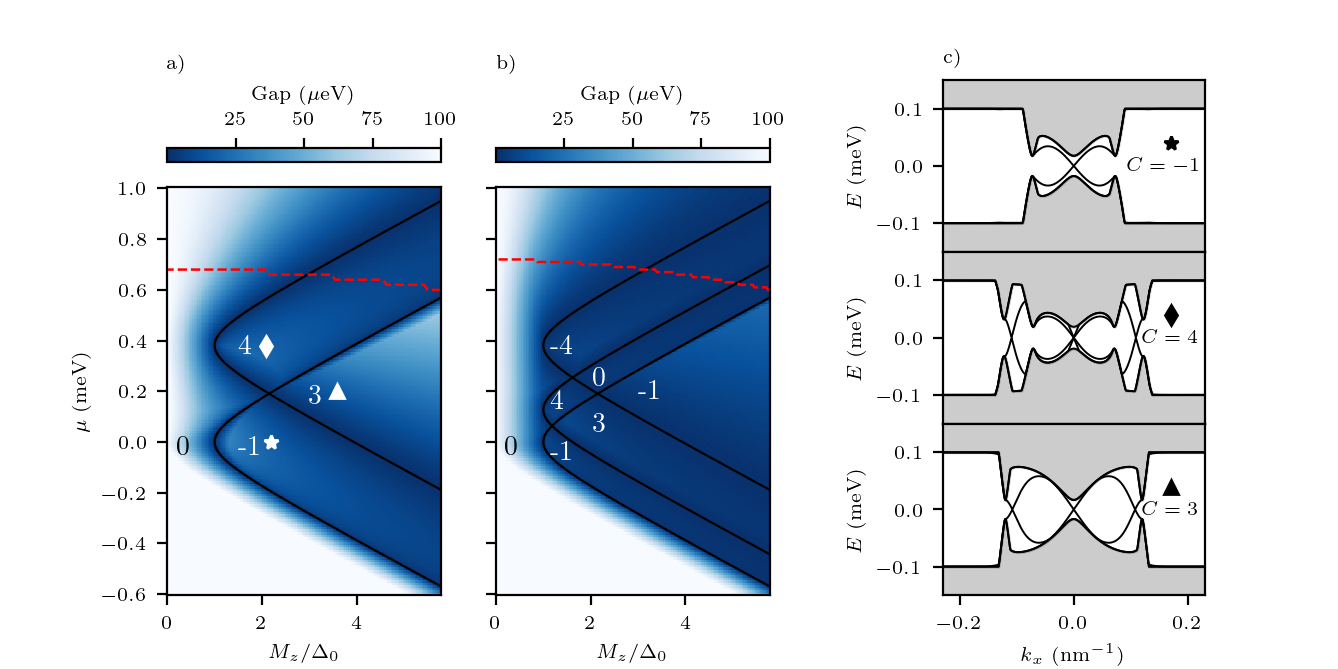}
    \caption{Spinful single-band model. We set $\Delta_0=100~\mu$eV throughout. (a) Topological phase diagram for $\alpha_1=0.03$ nm$^{-1}$, $\alpha_2=0.3$ nm and $\alpha_3=2.7$ nm. The solid line represents the critical hyperbolas that separate different topological phases, while the integers represent the corresponding value of the Chern number. The markers indicate the value of $M_z$ and $\mu$ used in panel (c). The red dashed line defines the region of validity of the model: below the red line the chemical potential is low enough to avoid the unphysical part of the energy dispersion of the bands at high momenta. (b) Topological phase diagram for representative values of the parameters $\alpha$ ($\alpha_1=0.01$ nm$^{-1}$, $\alpha_2=2$ nm and $\alpha_3=1$ nm).  Again, the red dashed line defines the region of validity of the model. (c) One-dimensional band structure of an infinite lead ($W = 9000$ nm) for the corresponding parameters values of the points marked in panel (a). Zero energy states are clearly visible, signaling the presence of edge states. For $C=4$ (middle panel), two right-moving edge states are present around $k_x=0$.}
    \label{fig:model_two}
\end{figure*}

\subsection{Superconducting properties and topological phase diagram}\label{sec:topo_model2}

Introducing the four-component Nambu spinor
\begin{equation}
\Psi_\mathbf{k}=\begin{bmatrix}
C_\mathbf{k} \\ C^\dagger_{-\mathbf{k}}\,,
\end{bmatrix}
\end{equation}
the Hamiltonian can be rewritten as
\begin{equation}
H = \frac{1}{2}\int d\mathbf{k}\,\Psi^\dagger_\mathbf{k}\mathcal{H}_\mathbf{k}\Psi_\mathbf{k}
\end{equation}
with the Bogoliubov-de Gennes Hamiltonian
\begin{equation}\label{eq:H_model2}
\mathcal{H}_\mathbf{k} = \begin{bmatrix}
\mathcal{h}_\mathbf{k} & i\Delta_0\sigma_y \\
-i\Delta_0\sigma_y & -\mathcal{h}^*_{-\mathbf{k}}
\end{bmatrix}\,.
\end{equation}
We want to identify distinct superconducting phases that may occur in this model upon varying, say, the chemical potential $\mu$ or the magnetization $\mathbf{M}$. Gap closing may be found by the condition $\det(\mathcal{H}_\mathbf{k})=0$.

If the magnetization is in-plane, we find that the gap closes when $\abs{\mathbf{M}}=\Delta_0$ and remains zero afterwards, signaling a transition into a phase of gapless superconductivity.
The transition occurs due to the tilting of the energy bands in momentum space and regardless of the orientation of $\mathbf{M}$ in the plane. Note that the anisotropy of the energy bands does not translate in an anisotropy of the critical magnetization, because the latter is due to a competition between the superconducting pairing and the magnetization, which in this model are both independent of momentum.

Topological phases can occur if the magnetization is out-of-plane, i.e. $\mathbf{M}=(0,0,M_z)$.
In this case, it is possible to find analytical expressions of the critical lines and the points in momentum space where the gap closes.
Indeed, the condition $\det(\mathcal{H}_\mathbf{k})=0$ yields 
\begin{equation}
\left(\zeta^2_\mathbf{k} - M^2_z+ \Delta_0^2 - |A_\mathbf{k}|^2\right)^2 + 4\Delta_0^2|A_\mathbf{k}|^2 = 0.
\end{equation}
To find all possible gap closing dictated by this equation, we again inspect all the points in momentum space such that $A_{\mathbf{k}} = 0$, which coincide with the spin-degeneracy points listed in the previous subsection. Then, the condition $\zeta_{\mathbf{k}}^2 -M_z^2 + \Delta_0^2 = 0$ gives the critical lines in the parameter space of the model, which are hyperbolas in the $(\mu, M_z)$ plane. 

We find the following solutions. First, the gap may close at $k=0$ along the critical curve
\begin{equation}\label{eq:transition_0}
\mu^2=M_z^2-\Delta_0^2
\end{equation}
At this gap closing a single Dirac cone is formed in momentum space, thus the Chern number changes by one. This is the same topological transition that occurs in Rashba 2DEGs.

Second, the gap may close simultaneously on the main diagonals when $k=k_0$. In this case the critical curve is
\begin{equation}\label{eq:transition_1}
\left(\mu-\frac{\hbar^2}{2m}\frac{\alpha_1}{\alpha_2+\alpha_3}\right)^2=M_z^2-\Delta_0^2\,.
\end{equation}
At this gap closing, four Dirac cones appear in momentum space, thus the Chern number changes by four.
If $\alpha_2>\alpha_3$, a similar four-fold gap closing can be found on the principal axes in momentum space when $k=k_1$, yielding the critical line
\begin{equation}\label{eq:transition_2}
\left(\mu-\frac{\hbar^2}{2m}\frac{\alpha_1}{\alpha_2-\alpha_3}\right)^2=M_z^2-\Delta_0^2\,.
\end{equation}
Again, the Chern number must change by four at this gap closing.
Finally, in the special case $\alpha_3=0$, the gap may also close on a circle of radius $(\alpha_1/\alpha_2)^{1/2}$ in momentum space. This is a trivial gap closing in the sense that the Chern number does not change, and will not play a role in the rest of the discussion.

Typical topological phase diagrams in the $(\mu, M_z)$ plane that result from the three transitions of Eqs.~\eqref{eq:transition_0},~\eqref{eq:transition_1},~\eqref{eq:transition_2} are represented in Fig.~\ref{fig:model_two}.
For each gapped region, we computed the Chern number numerically at representative points. The changes in Chern number across neighboring phases always coincide with the counting of Dirac cones at the critical lines separating them.

In panel~\ref{fig:model_two}a, we show a phase diagram for the case $\alpha_2<\alpha_3$.
There are two critical hyperbolae: the lower one corresponds to the gap closing at $k=0$, and the upper one corresponds to the gap closings at $k=k_0$.
Their overlap results in the presence of three distinct topological phases with $C=-1$, $C=4$ and $C=3$.
We note that reaching the phase with $C=3$ requires crossing both critical lines if one starts from the topologically trivial phase at $M_z=0$.
It is notable how the energy gap in the topological phase with $C=3$ grows with chemical potential: it is controlled by the spin-splitting of the energy bands, which grows with momentum and thus with chemical potential, unlike in a model with linear-in-momentum spin-orbit coupling where it saturates to a constant.

\begin{figure*}[t!]
    \centering
    \includegraphics[width=0.8\linewidth]{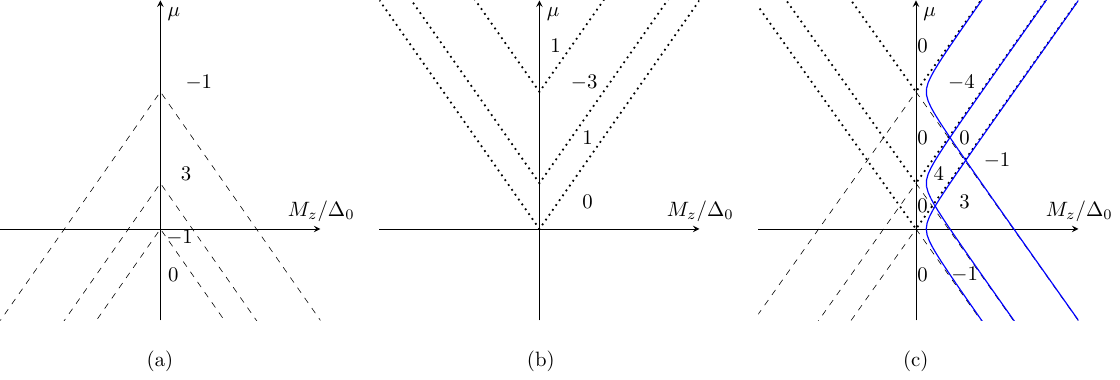}
\caption{Reduction of the topological phase diagram of the spinful single-band model as a sum of two independent copies of the spinless model of section \ref{sec:model1}. In blue there are the hyperbolas of the corresponding spinful single-band model.}
\label{fig:gap_diag_low+high}
\end{figure*}

In panel~\ref{fig:model_two}b, we show a phase diagram for the case $\alpha_2>\alpha_3$. 
In this case, there is a third critical hyperbola corresponding to the gap closing at $k=k_1$, leading to a new topological phase with $C=-4$.
The overlap of the three critical lines leads to a more complicated phase diagram at large $M_z$: in particular, with respect to panel (a), the phase with $C=3$ shrinks to leave room to a phase with $C=-1$ at large values of $M_z$.
Note how the two topological phase diagrams in panels (a) and (b) have distinct limits when $\alpha_1=0$: in both cases a single topological phase transition remains along the critical line~\eqref{eq:transition_0}, but the Chern number on the topological side is $C=3$ if $\alpha_2<\alpha_3$ and $C=-1$ if $\alpha_2>\alpha_3$.

In panel~\ref{fig:model_two}c, we show the numerically computed band structure of a translationally invariant lead along the $x$ direction, revealing the presence of chiral edge states and verifying the bulk boundary correspondence. We note in particular that in the $C=4$ phase there are two chiral edge states around $k=0$.

\subsection{Reduction to the spinless toy model}

Further intuition regarding the phase diagrams of Fig~\ref{fig:model_two}, and in particular about the occurrence of phases with $\abs{C}=4$, can be gained in terms of the spinless toy model discussed in the previous Section~\ref{sec:model1}.
Following Alicea~\cite{alicea2010}, we first introduce the normal modes which diagonalize the normal-state Hamiltonian $\mathcal{h}_\mathbf{k}$:
\begin{align}
d_{\mathbf{k},+}&=\cos(\beta_\mathbf{k}/2)\,c_{\mathbf{k}\uparrow} + e^{i\varphi_\mathbf{k}}\sin(\beta_\mathbf{k}/2)\,c_{\mathbf{k}\downarrow}\,, \\
d_{\mathbf{k},-}&=- e^{-i\varphi_\mathbf{k}}\sin(\beta_\mathbf{k}/2)\,c_{\mathbf{k}\uparrow} +\cos(\beta_\mathbf{k}/2)\,c_{\mathbf{k}\downarrow}\,,
\end{align}
with $\tan(\beta_\mathbf{k}) = \abs{A_\mathbf{k}}/M_z$ and $\varphi_\mathbf{k}$ the phase of $A_\mathbf{k}$.
Then, in terms of a new four-component Nambu spinor, $\Phi_\mathbf{k}~=~[d_{\mathbf{k},+}\;
d^\dagger_{-\mathbf{k},+}\;
d_{\mathbf{k},-}\;
d^\dagger_{-\mathbf{k},-}]^T$,
the Hamiltonian becomes $H=\tfrac{1}{2}\int d\mathbf{k} \Phi^\dagger_\mathbf{k} \tilde{\mathcal{H}}_\textrm{BdG}\Phi_\mathbf{k}$, with a rotated BdG Hamiltonian
\begin{equation}\label{eq:hbdg_tilde}
\tilde{\mathcal{H}}_\textrm{BdG}=
\begin{bmatrix}
E_{\mathbf{k},+} & \Delta_{\mathbf{k},++} & 0 & \Delta_{\mathbf{k},+-} \\
\Delta^*_{\mathbf{k},++} & -E_{\mathbf{k},+} & -\Delta_{\mathbf{k},+-} & 0 \\
0 & -\Delta_{\mathbf{k},+-} & E_{\mathbf{k},-} & \Delta^*_{\mathbf{k},++} \\
\Delta_{\mathbf{k},+-} & 0 & \Delta_{\mathbf{k},++} & - E_{\mathbf{k},-}
\end{bmatrix}\,.
\end{equation}
Here, $\Delta_{\mathbf{k},++}$ and $\Delta_{\mathbf{k},+-}$ are \emph{intra}-band and \emph{inter}-band pairing respectively:
\begin{align}
\Delta_{\mathbf{k},++} &=\frac{\Delta_0 A_\mathbf{k}}{\sqrt{M_z^2+\abs{A_\mathbf{k}}^2}}\,,\quad \Delta_{\mathbf{k},+-} =\frac{\Delta_0 M_z}{\sqrt{M_z^2+\abs{A_\mathbf{k}}^2}}\,.
\end{align}
In Eq.~\eqref{eq:hbdg_tilde}, note that the intra-band pairing for the upper and lower energy bands have opposite phases.
We also note that this transformation only works for the case $\mathbf{M}\parallel \hat{\mathbf{z}}$.

To reconstruct the topological phase diagram, we can focus first on each band separately, and then consider the effect of the inter-band pairing.
Because $\Delta_{\mathbf{k},++}$ is proportional to $A_\mathbf{k}$, when restricted to a single band the Hamiltonian takes the same form of the toy model of Sec.~\ref{sec:model1}, except with $E_{\mathbf{k},\pm}$ taking the place of $\zeta_\mathbf{k}$.

The resulting phase diagrams for upper and lower energy bands are schematically shown in Fig.~\ref{fig:gap_diag_low+high}(a) and (b), respectively.
Two features stand out from their comparison.
First, the phase transition lines have opposite slopes in the two panels, because they are dictated by the conditions $\zeta_\mathbf{k}=\mp \abs{M_z}$ respectively for the upper and lower energy bands.
Second, the Chern numbers are always opposites, because the intra-band pairing winds in opposite direction when circling the origin in momentum space.

To obtain the full topological phase diagram of the model we simply need to add the two panels (a) and (b), summing the Chern numbers of the two bands as shown in Fig.~\ref{fig:gap_diag_low+high}(c). Because of the property just mentioned, this results in the presence of phases with $\abs{C}=4$. The union of the critical lines coming from the two bands leads to the correct asymptotes of the critical lines of the full model. The effect of inter-band pairing $\Delta_{\mathbf{k},\pm}$ is to open up a gap when the energy bands are degenerate, as shown by the blue lines in panel (c). This reproduces all the features of the full phase diagram of the model.

\section{Effective single-band planar model}
\label{sec:model3}

In this Section we upgrade the single-band model to a more realistic variant, obtained from a self-consistent treatment of the confined hole gas.
Our goal is to describe the heterostructure sketched in Fig.~\ref{fig:sketch}.

Our starting point is the six-band Luttinger-Kohn Hamiltonian~\cite{luttinger1955,luttinger1956,bir-pik,terrazos2021} which gives a low-energy description of the valence bands of germanium: two heavy-hole bands ($j=\tfrac{3}{2}$, $m_j=\pm\tfrac{3}{2}$), two light-hole bands ($j=\tfrac{3}{2}$, $m_j=\pm\tfrac{1}{2}$), and the two split-off bands ($j=\tfrac{1}{2}$, $m_j=\pm\tfrac{1}{2}$). Ordering the basis as
\begin{equation}\label{eq:basis}
\left\{\ket{\tfrac{3}{2}, \tfrac{3}{2}}, \ket{\tfrac{3}{2}, \tfrac{1}{2}}, \ket{\tfrac{3}{2}, -\tfrac{1}{2}}, \ket{\tfrac{3}{2}, -\tfrac{3}{2}}, \ket{\tfrac{1}{2}, \tfrac{1}{2}}, \ket{\tfrac{1}{2}, -\tfrac{1}{2}}\right\}
\end{equation}
the Hamiltonian is:
\begin{widetext}
\begin{equation}
    \mathcal{H}_{\text{LK}} = 
    \begin{pmatrix}
    P + Q & -S & R & 0 & -S/\sqrt{2} & \sqrt{2}R \\
    -S^* & P - Q & 0 & R & -\sqrt{2}Q & \sqrt{3/2}S \\
    R^* & 0 & P - Q & S & \sqrt{3/2}S^* & \sqrt{2}Q \\
    0 & R^* & S^* & P + Q & -\sqrt{2}R^* & -S^*/\sqrt{2} \\
    -S^*/\sqrt{2} & -\sqrt{2}Q^* & \sqrt{3/2}S & -\sqrt{2}R & P + \Delta & 0 \\
    \sqrt{2}R^* & \sqrt{3/2}S^* & \sqrt{2}Q^* & -S/\sqrt{2} & 0 & P + \Delta
    \end{pmatrix}
\end{equation}
\end{widetext}
with
\begin{align}
P &= \frac{\hbar^2}{2m_0}\gamma_1(k_x^2 + k_y^2 + k_z^2)\,,\\
Q &= \frac{\hbar^2}{2m_0}\gamma_2(k_x^2 + k_y^2 - 2k_z^2)\,,\\
R &= \sqrt{3}\frac{\hbar^2}{2m_0}\left[-\gamma_2(k_x^2 - k_y^2) + 2i\gamma_3 k_xk_y\right]\,,\\
S &= \sqrt{3}\frac{\hbar^2}{m_0}\gamma_3(k_x - ik_y)k_z.
\end{align}
Here, $m_0$ is the bare electron mass, $\Delta=0.296$~eV is the energy gap to the split-off bands, and $\gamma_1=13.35, \gamma_2=4.25$ and $\gamma_3=5.69$ are the Luttinger-Kohn band parameters for germanium~\cite{winkler2003spin}.

As a source of time-reversal symmetry breaking, we consider the presence of dilute manganese impurities in germanium.
The exchange interaction between the $p$-orbitals of germanium valence states and the $d$-orbitals of manganese atoms leads to a magnetization density $\mathbf{m}=\abs{\mathbf{m}}\hat{\mathbf{w}}$ due to the ordering of the manganese spins~\cite{dietl2001}, either hole-mediated or induced by a small external magnetic field.
For an arbitrary orientation $\hat{\mathbf{w}}=(w_x, w_y, w_z)$ of the magnetization, the effect of the exchange interaction on the germanium holes is well described by the following Hamiltonian,
\begin{widetext}
\begin{equation}
\mathcal{H}_{pd} = M_0 x
\begin{pmatrix}
    3w_z & \sqrt{3}w_- & 0 & 0 & -\sqrt{6}w_- & 0 \\
    \sqrt{3}w_+ & w_z & 2w_- & 0 & 2\sqrt{2}w_z & -\sqrt{2}w_- \\
    0 & 2w_+ & -w_z & \sqrt{3}w_- & \sqrt{2}w_+ & 2\sqrt{2}w_z \\
    0 & 0 & \sqrt{3}w_+ & -3w_z & 0 & \sqrt{6}w_+ \\
    -\sqrt{6}w_+ & 2\sqrt{2}w_z & \sqrt{2}w_- & 0 & -w_z & -w_- \\
    0 & -\sqrt{2}w_+ & 2\sqrt{2}w_z & \sqrt{6}w_- & -w_+ & w_z
\end{pmatrix}.
\end{equation}
\end{widetext}
Here, $w_\pm = w_x\pm i w_y$ and $M_0\,x$ is the Zeeman energy coming from the exchange interaction, where $x$ is the Mn content and $M_0$ a constant whose magnitude is proportional to the exchange integral and the Mn magnetic moment.

Based on the parameters proposed in Ref.~\cite{dietl2001}, one can estimate $M_0 \approx  1~\textrm{eV}$~\footnote{Following Ref.~\cite{dietl2001} adapted to our notation, the constant $M_0$ is given by $M_0 = \abs{\beta} n \mu / (6 g \mu_B)$ where $n$ is the density of Ge atoms in bulk Ge, $\mu$ is the magnetic moment of a single impurity, and $\beta$ is the $p$-$d$ exchange integral. For a neutral Mn impurity, with its half-filled $d$ shell, $\mu=\tfrac{5}{2}g\mu_B$. The value of the exchange integral of a substitutional impurity of Mn in Ge is estimated in Ref.~\cite{dietl2001} based on the corresponding value in GaAs, which is $\beta N_0\approx -1.2$~eV where $N_0$ is the number of cation sites per unit volume in GaAs; see also Ref.~\cite{jungwirth2006}. Because $n = 2N_0$, we obtain in the end $M_0=\tfrac{5}{6}\abs{\beta} N_0 \approx 1~$eV.}. This estimate must be subject to reasonable skepticism because it is not material-specific, it was made with higher Mn concentrations in mind, it was not made for a planar system, and it does not directly take into account the holes' own exchange correlation energy.
Nevertheless, the analyses that will follow do not rely on a precise value of $M_0$, but on the fact that the exchange interaction sets an~$\approx eV$ energy scale.
This implies that very low concentrations of Mn atoms ($x \sim 0.001$) can be sufficient to provide a magnetization energy $M=M_0 x$ comparable to the superconducting gap~($\Delta_0\sim 0.1$~meV) and chemical potentials~($\mu\sim 1$~meV) of the proximitized hole gas.

The final ingredient of the model is the hole confinement. We consider a layer of unstrained planar germanium of thickness $L$ along the $z$ direction. The holes in germanium are subject to a confinement potential $V(z)$ describing a triangular quantum well next to the barrier-germanium interface at $z=0$.
Thus, the total Hamiltonian is
\begin{equation}
\mathcal{H}=\mathcal{H}_\textrm{LK}+\mathcal{H}_{pd}+V(z)\,.
\end{equation}
From this microscopic 3D Hamiltonian of the germanium system, we obtain an effective, planar, single-band Hamiltonian via a Schrieffer-Wolff transformation~\cite{SW,bir-pik}, as described next.

\subsection{Derivation of the effective planar model}

First, we set both the magnetization and the in-plane momentum $\mathbf{k}_\parallel=(k_x, k_y)$ to zero and solve the Schrödinger-Poisson problem along the $z$ direction, using the one-dimensional Hamiltonian
\begin{equation}
\mathcal{H}_0=\mathcal{H}_\textrm{LK}|_{\mathbf{k}_\parallel=\mathbf{0}}+V(z)\,.
\end{equation}
We choose the Fermi level to be at zero energy. For the potential $V(z)$, we use the boundary conditions
\begin{equation}
V(0)=\phi_0\,,\qquad V(L)=0.
\end{equation}
with $\phi_0 < 0$ the potential energy setting the depth of the triangular quantum well at the germanium-barrier interface.
For the accompanying Schrödinger problem, we require the wave function to vanish in $z=0$ and $z=L$.
We then solve the electrostatic problem self-consistently, diagonalizing $\mathcal{H}_0$ to find the charge density $\rho$ and then solving the Poisson equation to determine the potential $V(z)$. We repeat the calculation for different values of $\phi_0$, so that we are able to determine the range of $\phi_0$ such that a single spin-degenerate energy level of $H_0$ is occupied at zero temperature. Under this condition, the confined hole gas is purely two-dimensional, with a single spinful sub-band occupied. In the absence of strain, the occupied sub-band has a heavy-hole character due to the smaller zero-point energy of the heavy holes in the confinement potential $V(z)$.

\begin{figure*}[ht]
    \centering
    \includegraphics[width=0.9\textwidth]{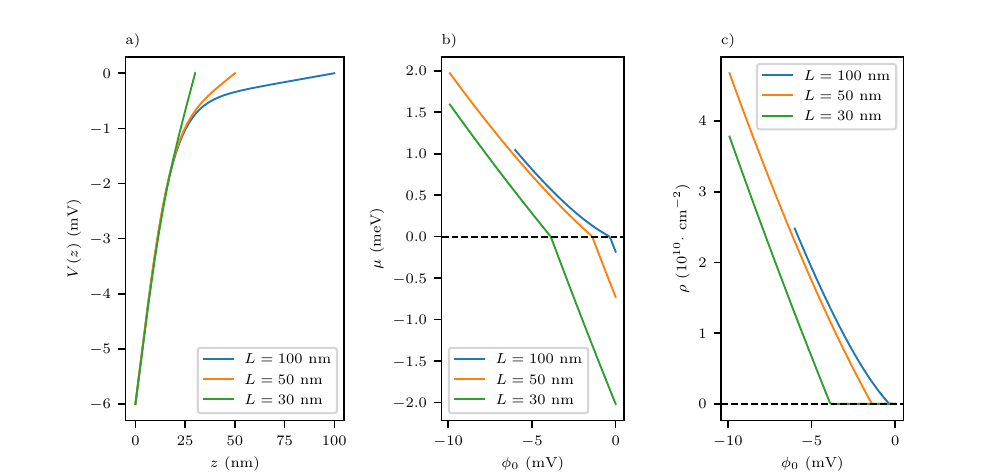}
    \caption{Effective single-band model. (a) Example of self-consistent potential $V(z)$ for different values of the thickness $L$. Chemical potential $\mu$ (b) and two dimensional density $\rho$ (c) as a function of $\phi_0$, for different values of the thickness $L$. For $L=100$ nm, the second band is occupied for $\phi_0<-6$ mV.}
    \label{fig:sp_results}
\end{figure*}

The second step is to perform a Schrieffer–Wolff transformation with the spin-degenerate ground state of $\mathcal{H}_0$ as the low-energy subspace and $\delta\mathcal{H}=\mathcal{H}-\mathcal{H}_0$ as the perturbation Hamiltonian.
The small parameters in the expansion are the magnetization $\mathbf{M}$ and the in-plane wave vector $\mathbf{k}_\parallel$.
Up to second order in the Schrieffer-Wolff transformation, the $2\times 2$ effective Hamiltonian has the form
\begin{subequations}\label{eq:heff}
\begin{align}\label{eq:h_eff_first_line}
\mathcal{h}_\mathbf{k} &= \tilde\zeta_\mathbf{k}\,\mathbb{I} + (\tilde{A}_\mathbf{k} \sigma_+ + \textrm{h.c.})  \\\label{eq:h_eff_second_line}
& +3 M w_z \sigma_z + M\left[\left(g_{\zeta}k_+^2 - g_{\bar{\gamma}}k_-^2\right) w_-\sigma_+ + \textrm{h.c.}\right]\\\label{eq:h_eff_third_line}
& + \left[\kappa_1 M (k_x w_y - k_y w_x) - \kappa_2 M^2(w_x^2+w_y^2)\right]\mathbb{I}
\end{align}
\end{subequations}
with $M=M_0 x$ and $k_\pm = k_x\pm i k_y$. The first term in Eq.~\eqref{eq:h_eff_first_line} is the effective hole kinetic energy,
\begin{equation}
\tilde{\zeta}_\mathbf{k} = \frac{\hbar^2k^2}{2m_0}\,\left[(\gamma_1 + \gamma_2 + A) + B\,k^2 + D\, \frac{(k_+^2-k_-^2)^2}{k^2}\right] -\mu\,,
\end{equation}
where $A, B$ and $D$ are output coefficients of the Schrieffer-Wolff transformation and the chemical potential $\mu$ is determined self-consistently from the Schrödinger-Poisson routine. The second term in Eq.~\eqref{eq:h_eff_first_line} is the spin-orbit interaction,
\begin{equation}
\tilde{A}_\mathbf{k} =  \frac{i\hbar^2}{2m}\left[-\alpha_2 k^2 (k_x+ik_y) + \alpha_3 (k_x-ik_y)^3\right]\,.
\end{equation}
With respect to the spin-orbit interaction used in Sec.~\ref{sec:model2}, we have omitted the term $\propto \alpha_1$, linear in momentum, since it does not occur for a heavy hole ground state.

In the second line of Eq.~\eqref{eq:heff} we gather the spin-splitting terms. Note in particular that the response to an in-plane magnetization couples spin and momentum via the non-zero coefficients $g_\zeta$ and $g_{\bar\gamma}$. Finally, the third line of Eq.~\eqref{eq:heff} gathers spin-independent corrections to the energy which are due to the magnetization. Again, we note that the in-plane magnetization couples directly to the hole momentum.

Overall, the effective Hamiltonian depends on nine coefficients that are obtained from the Schrieffer-Wolff transformation: $A, B$ and $D$; $\alpha_2$ and $\alpha_3$; $g_\zeta$ and $g_{\bar\gamma}$; and $\kappa_1$ and $\kappa_2$.
Explicit perturbative expressions for these parameters are given in Appendix~\ref{app:SW}. Explicit values of the parameters as a function of $\phi_0$ can be obtained from the energy levels and the wave functions of the excited states of $\mathcal{H}_0$, as computed in the final step of the Schrödinger-Poisson cycle.

\begin{figure*}[t]
    \centering
    \includegraphics[width=0.9\textwidth]{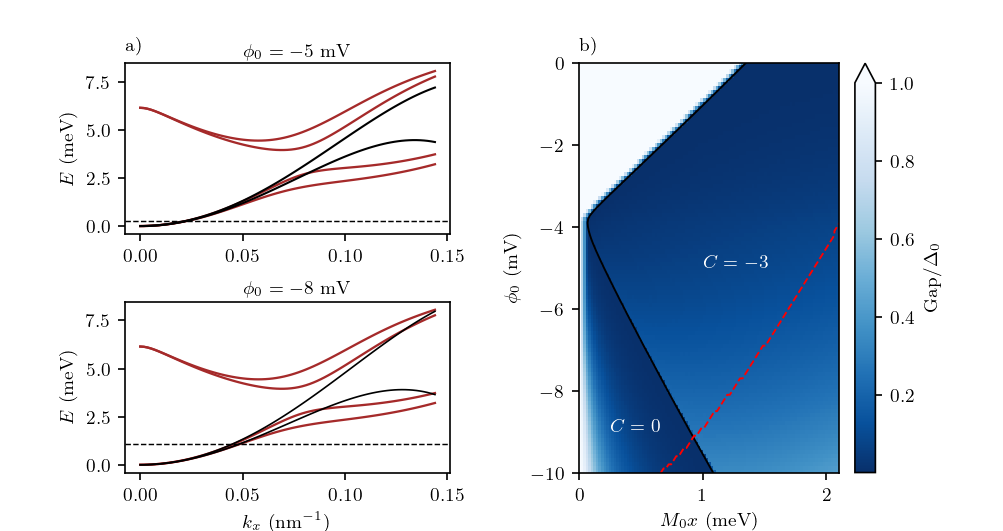}
    \caption{(a) Comparison between exact diagonalization of the confined Luttinger-Kohn Hamiltonian, in red, and the planar effective model, in black, for $\phi_0 = -5$ mV (upper panel) and $\phi_0 = -8$ mV (lower panel). The dashed line represents the chemical potential. (b) Topological phase diagram for unstrained Ge with out-of-plane magnetization $\mathbf{M}=M_0x\hat{\mathbf{e}}_z$ and induced gap $\Delta_0=0.1$~meV. The single-band planar model is not applicable below the red dashed line.}
    \label{fig:model3_out-of-plane}
\end{figure*}

The effective planar Hamiltonian is only valid at energies lower than the sub-band spacing that separates the ground state of $\mathcal{H}_0$ from the first excited state.
The failure of the effective model is signaled by the fact that the energy bands start bending down at large momenta, due to the cubic terms overcoming the quadratic terms, as already happened in the toy model of Sec.~\ref{sec:model2}.

As a final ingredient, $s$-wave superconductivity is introduced in the model in the same way as in the previous Section. Thus, the Bogoliubov-de Gennes Hamiltonian remains that of Eq.~\eqref{eq:H_model2} but featuring the effective Hamiltonian $\mathcal{h}_\mathbf{k}$ of Eq.~\eqref{eq:heff}.
We now present the numerical results obtained from this model.

\subsection{Numerical Results}

In Fig.~\ref{fig:sp_results} we show the typical results of the Schrödinger-Poisson calculation, for different values of the thickness $L$ of the germanium layer: $L=30$~nm, $L=50$~nm and $L=100$~nm.
In panel (a), we show the self-consistent potential $V(z)$ for a value \textcolor{black}{$\phi_0=- 6$~mV}.
The self-consistent potential profile consists of a triangular well with an effective width of approximately $20$~nm, after which the screening by the holes kicks into gear, as visible in particular for the $L=100$~nm calculation.

\begin{figure*}[t!]
    \centering
    \includegraphics[width=0.8\textwidth]{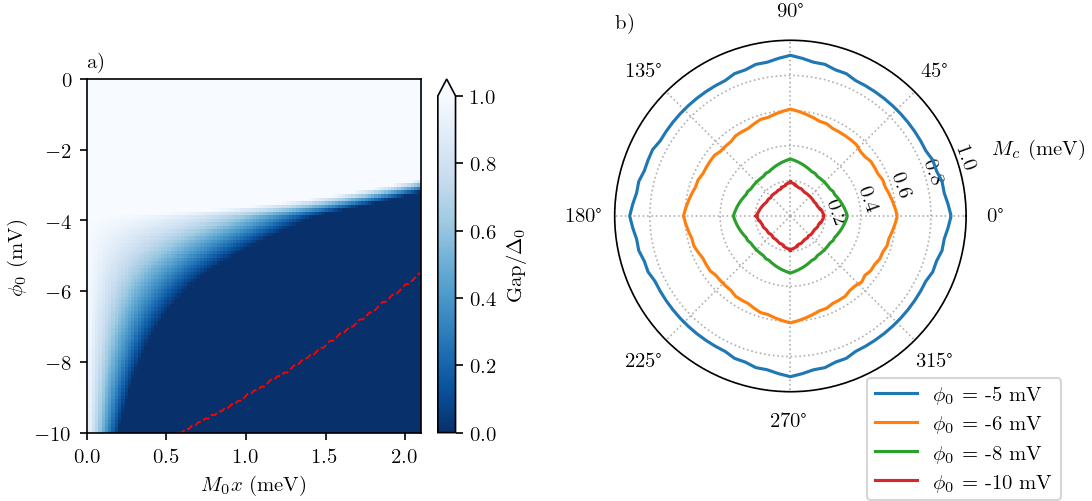}
    \caption{(a) Topological phase diagram for unstrained Ge with in-plane magnetization $\mathbf{M}=M_0\,x\,\hat{\mathbf{e}}_x$ and induced gap $\Delta_0=0.1$~meV. The gap closes and never reopens once the critical magnetization is reached. The single-band planar model is not applicable below the red dashed line. (b) Contour plot of the critical magnetization $\mathbf{M}_c$ for several values of $\phi_0$. For increasing magnitude of $\phi_0$ the anisotropy becomes stronger.}
    \label{fig:model3_compr_Mc}
\end{figure*}

In panels (b) and (c) we show the chemical potential and hole density obtained at the end of the self-consistent cycle, as a function of $\phi_0$. The range of $\phi_0$ is limited to the condition that a single sub-band is occupied. 
This constraint limits the calculation to $0 > \phi_0 > -10~$mV for $L=50$ and $30$~nm, and to $0 > \phi_0 > -6~$mV for the $L=100$~nm case, for which confinement effects are much weaker and the sub-band spacing correspondingly smaller.
For these relatively low values of $\phi_0$, the chemical potentials $\mu$ remain in the few meV range, as shown in Fig.~\ref{fig:sp_results}(b). The corresponding hole densities reach values of the order of $5\cdot 10^{10}$~cm$^{-2}$, as shown in Fig.~\ref{fig:sp_results}(c).
Overall, the results of our Schrödinger-Poisson calculations are consistent with the experimental picture reported by Costa et al.~\cite{costa2026} for an unstrained germanium channel.

\textcolor{black}{Note that in both panels (b) and (c) the effect of hard-wall confinement over the channel thickness $L$ are visible: the zero-point energy associated with a finite $L$ causes the chemical potential to be negative and the density to be zero when $\abs{\phi_0}$ is smaller than a threshold value that decreases with increasing $L$.}
From now on, all the results are obtained from simulations with $L=30$~nm.

In the Appendix~\ref{app:SW} we show the effective parameters computed for $L=30$~nm as a function of $\phi_0$.
In Fig.~\ref{fig:model3_out-of-plane}(a), we compare the band structure of the effective planar model (in black) with that of the confined Luttinger-Kohn model (in red), at zero magnetization and for two representative values of $\phi_0$.
The planar model captures reasonably well the curvature and cubic-in-momentum spin-splitting of the energy bands up until the Fermi level (horizontal dashed lines).
Consistent with expectations, it stops being an accurate low-energy description once the energy bands hybridize with the higher sub-bands present in the calculation based on the confined Luttinger-Kohn model.

We now move on to the analysis of the phase diagram of the effective planar model in the presence of superconducting pairing, for which the choice of the magnetization direction is the crucial factor.
As discussed in Ref.~\cite{dietl2001}, the interaction between germanium holes and $d$-orbitals of the Mn atoms is a source of magnetic anisotropy, implying the presence of an easy-axis of the magnetization with respect to the crystal lattice.
The easy axis depends on the hole density and so can change with chemical potential.
The analysis of Ref.~\cite{dietl2001} suggests that for an unstrained germanium well the easy-axis should be out-of-plane at low densities, and in-plane at higher densities.
This analysis, however, is carried out for a bulk system and in a regime of overall higher hole densities and manganese concentration than those of interest here, making it hard to extrapolate the results to our case.
Thus, we present results for both out-of-plane and in-plane magnetization.

In Fig.~\ref{fig:model3_out-of-plane}(b) we show the topological phase diagram for an out-of-plane magnetization, $\hat{\mathbf{w}}~=~\hat{\mathbf{e}}_z$, in the phase space formed by $\phi_0$ and the Mn concentration $x$.
As in Sec.~\ref{sec:model2}, we set $\Delta_0=100~\mu$eV.
The effective Hamiltonian for this case is very similar to that analyzed in Sec.~\ref{sec:model2}: so, the critical lines are still given by the roots of the spin-orbit interaction terms.
In particular, since the linear term with coefficient $\alpha_1$ is zero, the gap closes only in $k=0$, separating a trivial phase with $C=0$ at low concentrations from a topological phase with $C=3$ at higher concentrations $x$.

The critical magnetization at which the transition occurs increases -- roughly linearly -- with increasing density, thus for lower values of $\phi_0$.
As expected, the required impurity concentrations are rather low -- less than one percent -- because of the large difference between the $\sim$~eV energy scale associated with the exchange interaction and the $~0.1$~meV energy scale of the induced $s$-wave pairing.
The energy gap in the topological phases increases with hole density -- as already discussed within the toy model of Sec.~\ref{sec:model2} -- due to the fact that the spin-splitting of the energy bands increases with momentum.
The dashed red line defines the region of validity of the model, i.e. the region in the $(M_0x,\,\phi_0)$ plane where the chemical potential is always lower than the energies at the $k$-points where the bands of normal germanium bend downward.

In Fig.~\ref{fig:model3_compr_Mc}(a) we show the phase diagram for the in-plane magnetization along $x$, $\hat{\mathbf{w}}=\hat{\mathbf{e}}_x$.
As already pointed out in the study of the toy model in Section~\ref{sec:model2}, the in-plane magnetization closes the superconducting gap without reopening it, and so there are no topological phases in Fig.~\ref{fig:model3_compr_Mc}.
It is interesting to observe that the critical magnetization at which the gap closes depends strongly on density, and that it decreases with increasing density (lower values of $\phi_0$).
This is due to the momentum-dependence of the in-plane spin-splitting terms in Eq.~\eqref{eq:h_eff_second_line}, effectively leading to larger $g$-factors at higher hole densities.

A second interesting consequence of Eq.~\eqref{eq:h_eff_second_line} is that it leads to an anisotropic response, which depends on the direction of the magnetization in the plane.
This is verified in Fig.~\ref{fig:model3_compr_Mc}(b), which shows the critical magnetization leading to the closing of the superconducting gap as a function of the direction of $\hat{\mathbf{w}}$ in the plane, for a few values of $\phi_0$.
The anisotropy increases with density, in correspondence with smaller critical magnetizations, and it always retains the $C_4$ symmetry of the underlying model.

\section{Conclusions}

Using three models of increasing complexity, we have studied the possible occurrence of chiral topological phases in the proximitized hole gas of planar germanium.
Our calculations, performed in the regime of low densities, treat the confined Luttinger-Kohn model self-consistently as the starting point for a second-order Schrieffer-Wolff transformation leading to an effective planar model.
Building on the results of Ref.~\cite{mao2011}, we find that in unstrained germanium a topological phase with $C=3$ may occur provided that time-reversal symmetry is broken by a spin-splitting field in the out-of-plane direction.

In our effective planar model, there is no linear-in-momentum spin-orbit interaction term: it vanishes because within the confined six-band Luttinger-Kohn Hamiltonian there is no mixing between heavy and light holes at zero momentum.
This is consistent with previous findings in a similar framework~\cite{Marcellina2017}.
A linear term may be present, for instance, if microscopically the $z=0$ interface reduces the symmetry from $C_{4v}$ to $C_{2v}$~\cite{durnev2014}.
For similar reasons, the linear term could be sizeable if the heterostructure is grown along the $[110]$ direction~\cite{xiong2021}.
As the toy-model topological phase diagrams of Secs.~\ref{sec:model1} and~\ref{sec:model2} show, if a linear term is present, then topological phases with $\abs{C}=1$ or $\abs{C}=4$ may also occur.

As a concrete proposal that does not compromise the coexistence with the superconducting proximity effect from an $s$-wave superconductor deposited on top of planar germanium, we have considered the magnetization of dilute manganese impurities in germanium, a platform well known from the study of dilute ferromagnetic semiconductors~\cite{dietl2001,jungwirth2006,dietl2014}.

In principle, an ensemble of very dilute manganese atoms -- with Mn content of order $0.1\%$ -- should be sufficient to provide a net magnetization density large enough to drive the system into the topological phase.
This conclusion should be corroborated by a more in-depth modeling of hole-mediated ferromagnetism in the regime of very dilute impurities and low hole densities.
In particular, the self-consistent calculation of the magnetic easy-axis in the planar geometry and at low densities is an interesting question for future research.
If the critical temperature for ferromagnetic ordering turns out to be too low compared to the dilution refrigerator temperatures $\sim 50$~mK, one may rely instead on the paramagnetic response of the Mn spins to order them via a small perpendicular magnetic field.

Another important extension of our analysis concerns the role of strain.
In this work we have limited ourselves to unstrained germanium quantum wells, which have been recently proposed to enhance the spin-orbit interaction in planar germanium~\cite{costa2026}.
Compressive strain suppresses the spin-orbit interaction of heavy holes~\cite{bosco2021,coppini2026}: thus, it would presumably lead to qualitatively similar results as in the unstrained case, but with smaller gaps in the topological phase.
On the other hand, the case of tensile strain is more complex: it leads to a ground state of light holes~\cite{delvecchio2024,delvecchio2025}, for which the spin-orbit interaction would likely have $\alpha_2>\alpha_3$ as well as a linear term~\cite{coppini2026}.
This would change the topological phase diagram to something more similar to that of Fig.~\ref{fig:model_two}(b).
Our model of Sec.~\ref{sec:model3}, based on the six-band Luttinger-Kohn Hamiltonian, is well suited for an extension to the strained case, which we leave to future work.

Setting aside the topological phase, the regime of gapless superconductivity reached via an in-plane magnetization could also be interesting in itself.
In the clean limit, it is characterized by the anisotropy of the magnetic response and the presence of Bogoliubov-Fermi surfaces~\cite{agertberg2017,yuan2018,phan2022}.

Finally, let us return to the issue of disorder.
Breaking time-reversal symmetry via an ensemble of manganese impurities -- even if neutral -- is certain to lead to the presence of disorder, to the detriment of the high mobility of the hole gas in pristine germanium.
We may estimate a target mobility using the rough criterion that the scattering time $\tau$ should be larger than $\hbar/\Delta_\textrm{top}$, where $\Delta_\textrm{top}$ is the gap in the topological phase, and so that $\mu$ should be larger than $e\hbar/m\Delta_\textrm{top}$, where $m$ is the effective mass of the heavy holes.
Then, using $m = m_0/(\gamma_1+\gamma_2+A)$ and $\Delta \approx 0.1~$meV as an optimistic upper bound for the topological gap, the mobility should at least be larger than $\approx 3\cdot 10^5$~cm$^2/(\textrm{V}\cdot\textrm{s})$.
To this estimate we add that, while the predicted values of required impurity concentrations in our calculations are low, the hole gas is itself in a low-density regime, and so may be particularly prone to localization.
The effect of disorder would be to decrease the topological gap and possibly drive the system out of a gapped topological phase and into the so-called thermal metal phase~\cite{senthil2000,chalker2001}, whose phenomenology may be equally interesting to investigate experimentally.

Even with these caveats, the topological phase diagrams obtained here in the clean limit can be useful as guidance and motivation for the design and realization of novel heterostructures to explore the topological phase diagram of class $D$ in two dimensions.
In the meantime, a theoretical investigation of both strain and disorder in the model, complemented by quantum transport simulations, is a natural extension of the analysis presented in this work.

\subsection*{Data availability}

The code and data that support our findings will be made available before publication.

\subsection*{AI statement}

During the final stages of this work, we have used Claude Sonnet 5.5 for debugging code and checking the mathematical equations obtained analytically for the Schrieffer-Wolff coefficients in Appendix~\ref{app:SW}. We have also used Claude Opus 5.5 to review the quality of the writing in the final manuscript and its overall clarity.

\begin{acknowledgments}
VC is grateful for the hospitality of QuTech and TU Delft while carrying out part of the work. In particular, we are grateful to Anton Akhmerov, Michael Wimmer and the Quantum Tinkerers group for sharing computing resources with us. We thank Anton Akhmerov also for feedback on the manuscript. VC was supported by European Union - Next Generation EU,  Mission: I.3.4 Borse PNRR transizioni digitali e ambientali (Missione 4),  Component: 1, CUP: B53C23001600006.
\end{acknowledgments}

\appendix
\section{Details on the Schrieffer-Wolff coefficients}
\label{app:SW}

Because $\mathcal{H}_0$ is block diagonal in the basis~\eqref{eq:basis}, its eigenvectors can be divided into three groups.
The first group consists of eigenstates with a purely heavy-hole character:
\begin{align}
\ket{H_n,\pm} &= \ket{H_n}\otimes\ket{\tfrac{3}{2},\pm\tfrac{3}{2}}\,.
\end{align}
Here $n=1,2,3,\dots$ labels the solutions in order of increasing energy ($n=1$ is the ground state), while $\ket{H_n}$ denotes the spatial part of the solution, with wave function $H_n(z) = \braket{z}{H_n}$.

The second group consists of eigenstates which, at least at energies below the gap to the split-off band, $\Delta=0.296$~eV, have a predominantly light-hole character but also non-zero split-off hole component, due to the term $Q|_{\mathbf{k}_\parallel=0}$ which mixes them in $\mathcal{H}_0$.
These are:
\begin{align}
\ket{L_n,+} &= \ket{L_n}\otimes\ket{\tfrac{3}{2},+\tfrac{1}{2}} + \ket{\textrm{SO}_n}\otimes\ket{\tfrac{1}{2},+\tfrac{1}{2}}\,,\\
\ket{L_n,-} &= \ket{L_n}\otimes\ket{\tfrac{3}{2},-\tfrac{1}{2}}-\ket{\textrm{SO}_n}\otimes\ket{\tfrac{1}{2},-\tfrac{1}{2}}\,.
\end{align}
Here, $\ket{L_n}$ and $\ket{SO_n}$ are the two spatial wave functions for the light-hole and split-off hole components of the eigenstate. Note that they are not individually normalized to one.

Importantly, the eigenvalues of $\mathcal{H}_0$ do not depend on the quantum number $m_j$ and therefore we deal with two sets of spin-degenerate energy eigenvalues: $E_{H,n}$ for the heavy-hole states $\ket{H_n,\pm}$ and $E_{L,n}$ for the predominantly light-hole states $\ket{L_n,\pm}$.

\begin{figure*}
    \centering
    \includegraphics[width=\textwidth]{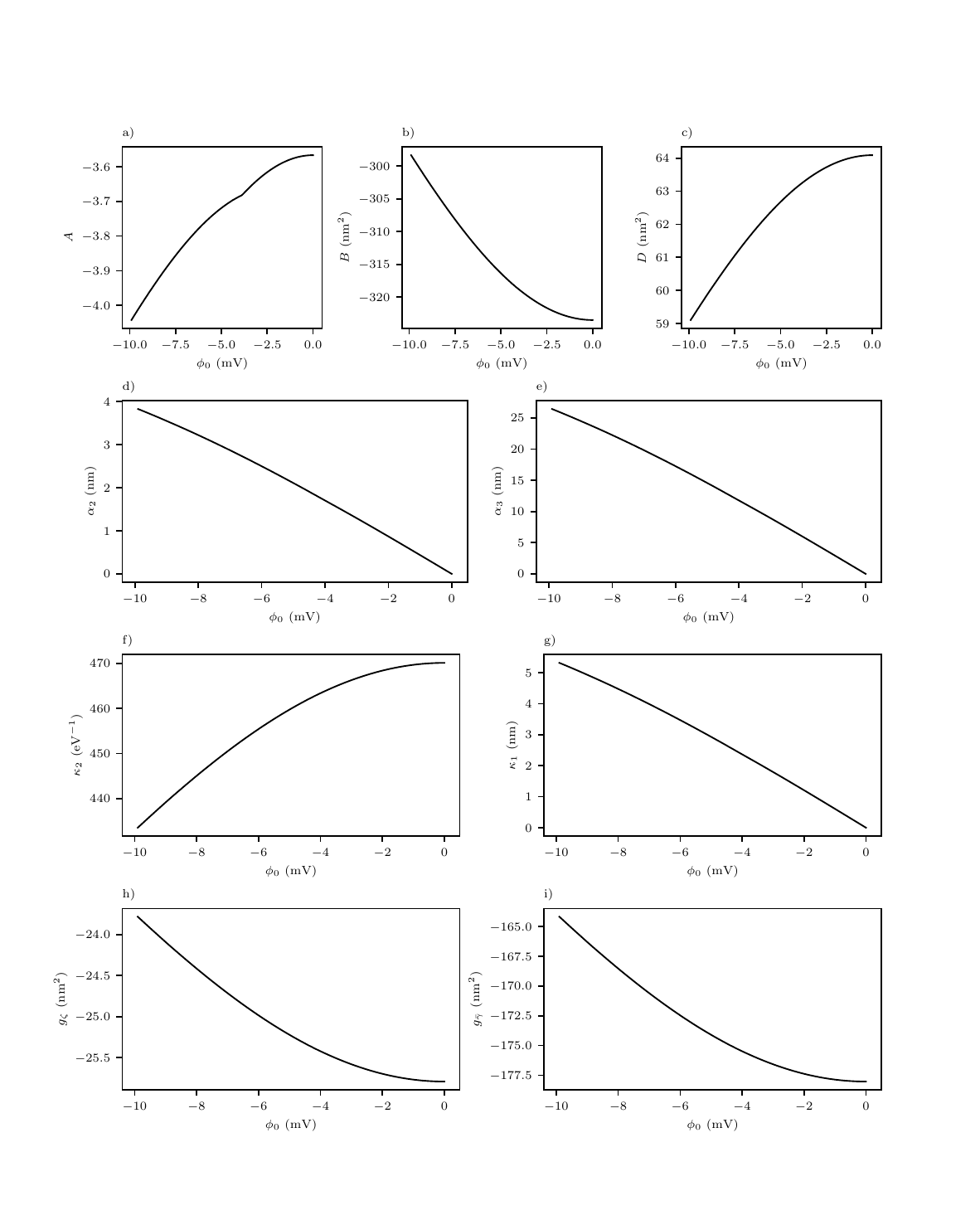}
    \caption{Schrieffer-Wolff coefficients. (a) $A$, (b) $B$, (c) $D$, (d) $\alpha_2$, (e) $\alpha_3$, (f) $\kappa_2$, (g) $\kappa_1$, (h) $g_\zeta$, (i) $g_{\bar{\gamma}}$ as a function of $\phi_0$.}
    \label{fig:Fig_model3_all_coeff}
\end{figure*}

Now that we have set the notation, we can give the results of the second-order Schrieffer-Wolff calculations. We start with the parameters $A$, $B$, $D$ which are
 are the corrections to the usual quadratic dispersion:
 
\begin{widetext}
    \begin{align}
    A &= 3\frac{\hbar^2}{2m_0}
    \gamma_3^2\sum_n\frac{|2\langle H_1|k_z\ket{L_n} +\sqrt{2}\langle H_1|k_z\ket{\textrm{SO}_n}|^2}{E_{H,1} - E_{L,n}}, \label{eq:A_coeff}\\
    B &= 3\frac{\hbar^2}{2m_0}(\bar{\gamma}-\zeta)^2\sum_n \frac{|\braket{H_1}{L_n} -\sqrt{2}\langle H_1\ket{\textrm{SO}_n}|^2}{E_{H,1} - E_{L,n}}, \\
    D &= -3\frac{\hbar^2}{2m_0}\bar{\gamma}\zeta\sum_n \frac{|\braket{H_1}{L_n} -\sqrt{2}\langle H_1\ket{\textrm{SO}_n}|^2}{E_{H,1} - E_{L,n}}.
    \end{align}
with $\bar\gamma = (\gamma_2+\gamma_3)/2$ and $\zeta=(\gamma_3-\gamma_2)/2$. The spin-orbit interaction coefficients are
\begin{align}
    \alpha_2 &= -3\frac{\hbar^2}{2m_0}\zeta\gamma_3\sum_n 2\Im\left\{ \frac{\left(\braket{H_1}{L_n} -\sqrt{2}\braket{H_1}{\textrm{SO}_n}\right)\left(2\bra{L_n} k_z \ket{H_1} +\sqrt{2}\bra{\textrm{SO}_n} k_z \ket{H_1}\right)}{E_{H,1} - E_{L,n}}\right\}, \\
    \alpha_3 &= 3\frac{\hbar^2}{2m_0}\bar{\gamma}\gamma_3\sum_n 2\Im\left\{\frac{\left(\braket{H_1}{L_n} -\sqrt{2}\braket{H_1}{\textrm{SO}_n}\right)\left(2\bra{L_n} k_z \ket{H_1} +\sqrt{2}\bra{\textrm{SO}_n} k_z \ket{H_1}\right)}{E_{H,1} - E_{L,n}}\right\}.
\end{align}
Finally, $\kappa_2$, $\kappa_1$, $g_\zeta$ and $g_{\bar{\gamma}}$ are the magnetic coefficients:
\begin{align}
    \kappa_1 &= -6i\frac{\hbar^2}{2m_0}\gamma_3\sum_n \frac{\left(\braket{H_1}{L_n} -\sqrt{2}\langle H_1\ket{\textrm{SO}_n}\right)\left(2\bra{L_n} k_z \ket{H_1} +\sqrt{2}\bra{\textrm{SO}_n} k_z \ket{H_1}\right)}{E_{H,1} - E_{L,n}}, \\
    g_{\zeta} &= \frac{3}{2}\frac{\hbar^2}{2m_0}\zeta\sum_n 2\Re\left\{\frac{\left(\braket{H_1}{L_n} - \sqrt{2}\langle H_1\ket{\textrm{SO}_n}\right)\left(-\braket{L_n}{H_1} + \sqrt{2}\braket{\textrm{SO}_n}{H_1}\right)}{E_{H,1} - E_{L,n}}\right\}, \\
    g_{\bar{\gamma}} &= \frac{3}{2}\frac{\hbar^2}{2m_0}\bar{\gamma}\sum_n 2\Re\left\{\frac{\left(\braket{H_1}{L_n} - \sqrt{2}\langle H_1\ket{\textrm{SO}_n}\right)\left(-\braket{L_n}{H_1} + \sqrt{2}\braket{\textrm{SO}_n}{H_1}\right)}{E_{H,1} - E_{L,n}}\right\}, \\
    \kappa_2 &= -3\sum_n \frac{|\braket{H_1}{L_n} - \sqrt{2}\langle H_1\ket{\textrm{SO}_n}|^2}{E_{H,1} - E_{L,n}} \label{eq:gz_coeff}
\end{align}
\end{widetext}

\clearpage
\bibliography{references}

\end{document}